# The significance of Halley's Comet in the Bayeux Tapestry

Michael Lewis & Simon Portegies Zwart

A comet appears in the Bayeux Tapestry between the scene showing the death of the English king Edward 'the Confessor' and the 'election' of his successor, Harold 'Godwinson' (Figure 1). The Tapestry's inscription only refers to this as a "star", though we can see from its depiction – shown with a 'hairy tail' – that it is a comet, now known to us as Halley's Comet (P1/Halley), after the English astronomer Edmond Halley who in 1682 established its average 76-year cycle of appearance in the Earth's skies.

Halley reported on his discovery to Isaac Newton, aware that, due to a myriad of internal and external processes, including the chaotic nature of its orbit,[1] its cycle, though regular, was not precise. Eighteen years after the publication of Newton's (1687) *Principia*, Halley also realised that this comet could be seen multiple times throughout history[2] and that it coincided with important happenings.[3] As we will show, he was not the first to observe this.

The depiction of Halley's Comet in the Bayeux Tapestry is a reminder of the importance of celestial events for people living around the year 1000. Also discussed below are two more comets (C/905 K1 and X/975 P1) identified with disasters that struck the English around the same time as the 1066 comet, and a further two (C/1097 T1 and X/1097 X1) that signify bad omens; the only positively perceived object associated with a comet (in 995) does not appear in modern records.

In historical accounts of many cultures, comets are generally considered portents of change rather than disaster.[4] Here we consider the significance of the Tapestry's comet in the context of the (so-called) English 'succession crisis' of 1066 with reference to (these) other contemporary accounts of comets. We conclude that although the tapestry's illustration is suggestive and unmistakably Halley's Comet, it is not a priority for its creators to give a precise account of its arrival in the sky, but rather connect it – likely for political reasons, albeit retrospectively – to the sequential events of Edward's death and Harold's coronation. In that, the tapestry's anonymous 'artists' provide a unique telling of its arrival.

---

1     Boekholt *et al.* 2016.
2     Newton 1695.
3     Halley 1705.
4     Seargent 2009.

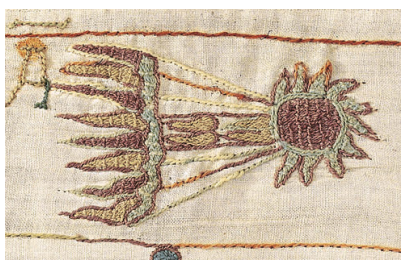

Figure 1: Detail of Halley's Comet. Bayeux Tapestry, scene 32. Image: Bayeux Museum.

**The 'succession crisis' of 1066**

The fact that Edward became King of England in 1042 is somewhat remarkable, given that he had lived almost 30 years of his life in exile. In 1013, Danes invaded England, resulting in Edward's father, King Æthelred II, sending his family abroad to Normandy.[5] Here, Edward and his siblings were brought up at the Norman court under the protection of Duke Richard II – brother of Edward's mother and Æthelred's wife, Queen Emma.[6] Æthelred continued to resist the invaders, but upon his death (23 April 1016), followed by that of his eldest son, Edmund 'Ironside' (30 November), Cnut 'the Great' succeeded to the whole kingdom. In a surprising twist, Emma returned to England to marry her husband's conqueror, leaving her children in Normandy.

Emma's union with Cnut resulted in a son, Harthacnut (born c.1018).[7] About this time, Cnut also succeeded to the Kingdom of Denmark and, by 1028, was also ruling Norway. However, upon Cnut's death (12 November 1035), this Anglo-Scandinavian Empire was to fall apart. Although Harthacnut was in Denmark, he could not prevent Magnus Olafsson from assuming power in Norway. This meant that even though Emma supported the succession of Harthacnut (over her sons by Æthelred), she was unable to prevent Harold 'Harefoot' (Cnut's son by Ælfgifu 'of Northampton') from becoming king. With her influence diminishing, Emma rekindled her relationship with her sons by Æthelred, who, in 1036, she encouraged to come to England with a military force. This mission ended in failure. Alfred was betrayed, captured and blinded, dying from his wounds, a crime levelled at the influential Anglo-Danish earl, Godwin.[8] Emma fled to Flanders to be with her daughter, Godgifu.

Emma's fate was to change favourably with Harold I's death (17 March 1040). Again, she backed Harthacnut to take the English throne, both triumphantly returning to England on 17 June. However, Harthacnut's autocratic rule proved unpopular, so (in 1041) he invited Edward to join him as co-ruler and heir.[9] Suddenly, on 8 June 1042, aged just 23-24, Harthacnut died, returning England to Æthelred's bloodline. Early in his kingship (1045), Edward married Edith 'Godwinson', the daughter of the aforementioned Earl Godwin 'of Wessex'. Given that Edward was suspicious of Godwin's role in the murder of his brother, Alfred, this was clearly a political choice designed to unite the bloodlines (and ambitions) of both men. Unfortunately, Edward and Edith's marriage was fruitless, adding to the tensions between the King and Earl. Things reached a head in September 1051 when Godwin was banished, and his family fled into exile.[10]

Godwin's absence enabled Edward to promote 'foreigners' to high office. It might also have been at this time that Edward mooted the idea that his cousin, Duke William of Normandy, might succeed him as king.[11] The situation again suddenly changed the following year when Godwin returned to England with significant armed forces, resulting in him being "granted his earldom as fully and as completely as he ever owned it, and his sons all just what they earlier owned, and his wife and daughter as fully and as completely as they earlier owned".[12] Moreover, the 'foreigners' at court were expunged. In such a climate of total capitulation, it was fortunate for Edward that Godwin died (15 April 1053) and was succeeded (in Wessex) by his second-eldest son, Harold, whom the King grew to respect and trust.[13]

Significantly in terms of the English succession, it had been learnt that a son of Edmund 'Ironside' was alive and well at the Hungarian royal court, and plans were made to return this 'Ætheling' (one who is throne-worthy) to England.[14] In 1057, Prince Edward set foot on English soil for the first time in over 40 years – but died before meeting the King. With him was his six-year-old son, Edgar, who was brought up at the royal court, with the presumption being that he would one day become king.[15] The English succession crisis was resolved, or so it seemed…

By late 1065, Edward was gravely ill, too ill, in fact, to attend the consecration of his new foundation, Westminster Abbey, on 28 December. Bedridden, the King was coming in and out of consciousness, and his mutterings both confused and scared those present.[16] In this state, Edward asks Harold to 'protect' Edith and the Kingdom, no doubt realising that Edgar needed help in defending England from likely invasions. But upon Edward's death (5 January 1066), the Witan (the royal council) elected Harold as king. He was crowned Harold II the very next day.[17]

---

5   Lavelle 2002, 129; Ashe & Ward 2020.
6   Stafford 2001, 223-224.
7   Howard 2008.
8   Barlow 2002, 30-31.
9   Licence 2020, 128-143.
10  Barlow 2002, 39-44.
11  Bates 2016, 108-119.
12  *ASC* (C), 180.
13  Walker 1997.
14  Ronay 1989.
15  Hooper 1985; Licence 2020, 227-232.
16  *VÆ*, 74-77.
17  Musgrove & Lewis, 201-203.

## The succession crisis and the Bayeux Tapestry

The Bayeux Tapestry conveys well the speed of Harold's election (scene 29) and consecration (scene 30) following the death of King Edward (Figure 2), but significantly it also links Harold's coronation (scene 31) to Halley's Comet being seen in the sky.[18] The men witnessing Harold's assent almost spin around towards the next scene (32) to point at the sky towards a "star" with a fiery tail. The tapestry's inscription says that "these men marvel at the star" (ISTI MIRANT STELLA[M]) – its terse Latin text being typical of the work, leaving much open to interpretation.

It is unknown when the tapestry was made, where, by whom and for what reason. Still, there is a general consensus that it was created between 1072 and 1077, probably in Canterbury, likely under the patronage of William of Normandy's half-brother, Bishop Odo of Bayeux.[19] If Odo was the patron, then his motivation for having it made might have been to highlight (even exaggerate) his role in the Norman Conquest of England. Although somewhat conjectural, it is even possible that it was made to tour Odo's earldom, perhaps being exhibited in castles and great halls. The tapestry's eventual home was Odo's cathedral of Bayeux, and it was perhaps even made for its consecration on 14 July 1077.[20] It might also be the case that the motivation for the project (from touring piece to church exhibit) changed considering a new political environment in the 1070s, whereby William, now king, lost patience with his increasingly perfidious and openly rebelling English subjects.[21]

Some have viewed the Bayeux Tapestry as purely Norman propaganda,[22] but a closer inspection shows it to be a complex work. Its message is subtle, even ambiguous, and this is likely to be purposeful.[23] It is almost certain that between 1066 and 1070, the Normans hoped the English would accept the new regime, and it was probably within this environment that the tapestry was conceived.[24] Furthermore, it tells a binary tale of two men fighting for control of England, completely ignoring the claims of others – notably Harold 'Hardrada' of Norway, who also invaded England in 1066, and Edgar 'Ætheling', who was still very much active at the time the tapestry was being made.[25] Indeed, it is the tapestry's extraordinary retelling of the events leading to the Norman Conquest of England in 1066 and its representation of the 'succession crisis' that makes its rendition of Halley's Comet so intriguing.

The Bayeux Tapestry begins its story in 1064 or 1065. The first scene (1) shows Edward and two men in conversation, one presumably Harold 'Godwinson', for thereafter (scene 2), Harold – who is now named in the inscription – makes for the south coast. At Bosham (scene 3), Harold and a companion are shown praying at the local church before the English feast (scene 4), thereafter embarking ships out to sea (scene 5). This journey sees them cross the English Channel and arrive in the lands of Count Guy of Ponthieu (scene 6), a vassal of Duke William. Here Harold is forcefully detained (scene 7) and brought back to Guy's castle at Beaurain (scene 8-9) to be interrogated.

Norman sources are clear that the reason for Harold's voyage was to confirm Edward's promise of the English crown,[26] though some later sources give alternative explanations. One is that Harold was looking to free English hostages held at the Norman court, including his brother Wulfnoth and nephew Hakon.[27] Or, less likely, Harold was on a fishing trip and was blown off course.[28] Instead, it seems likely that Harold intended to meet with William, but ended up in Ponthieu by mistake, perhaps even being shipwrecked there.

Next, the tapestry reverses events to show William learning of Harold's arrival (scene 12), his messengers making for Guy's castle to retrieve him (scene 11), and the said messengers in conversation with Guy (scene 10). Finally, Guy brings Harold to William (scene 13), and here ends the first length of the tapestry. Harold is then brought to William's palace, presumable Rouen, where the Duke and Earl talk (scene 14) (Figure 3). Shown next to Harold is a bearded individual, sometimes interpreted as one of the hostages Harold sought to release.[29] Also illustrated is the enigmatic scene (15) of Ælfgyva with a priest. This seems significant to the conversation between Harold and William, though it is not known for sure who the woman is or why her liaison with a priest is relevant to the tapestry's story.[30] Important for us is that much is open to interpretation, just like the depiction of Halley's Comet that comes later.

Harold then joins William on a military expedition against the Duke of Brittany, Conan II, in support of the rebel lord Rivallon of Dol. He is shown rescuing Normans from the quicksand of the River Couesnon (scene 17) as William's army crosses the border between Normandy and Brittany. The Normans then lay siege to Dol (scene 18), pass by Rennes and then attack Dinan (scene 19). Here, Conan surrenders to William (scene 20) and William 'gives arms' to Harold (scene 21). The fact that the Norman chronicler William of Poitiers gives quite a different version of these events is unimportant for us, apart from highlighting how

---

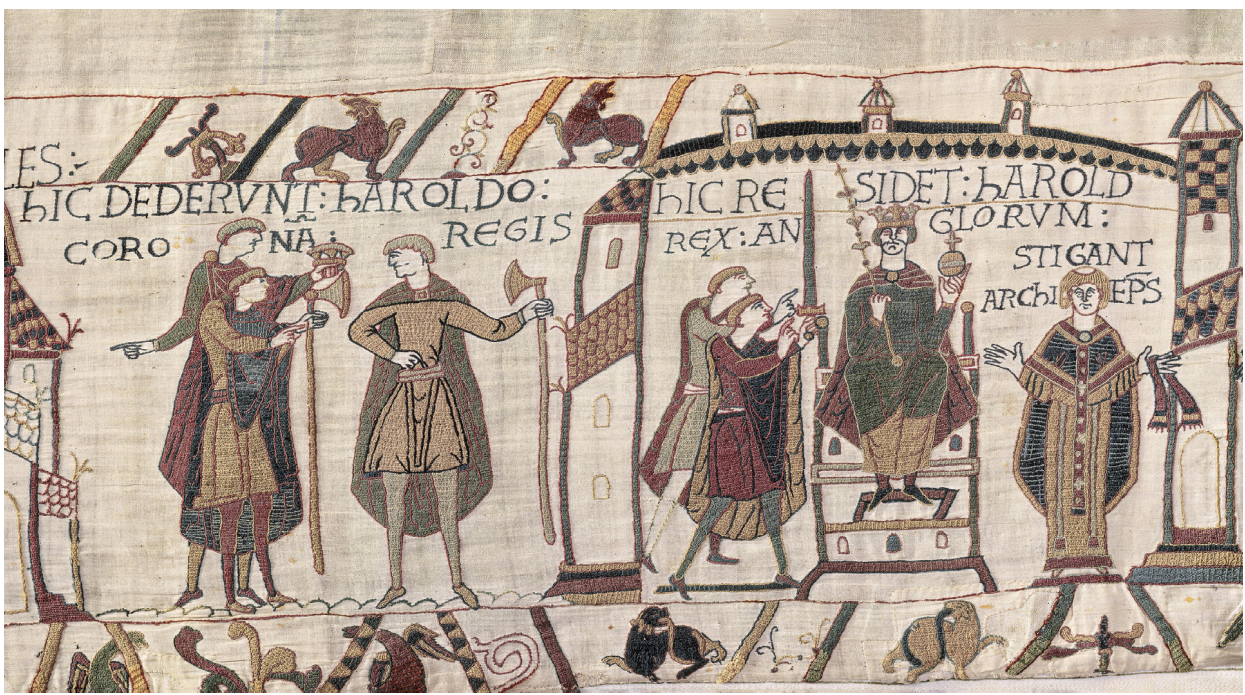

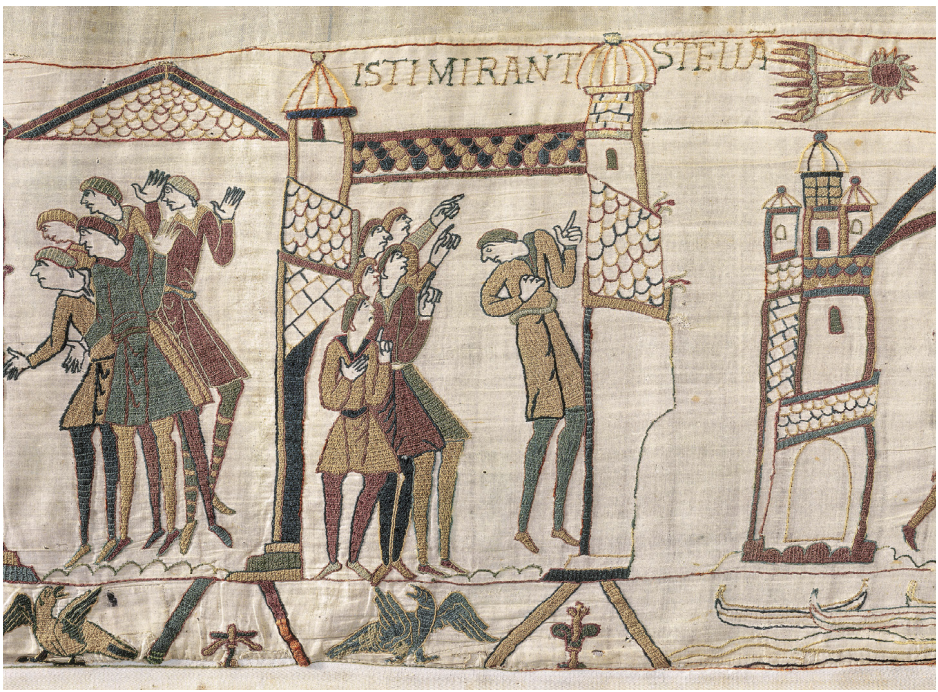

Figure 2: Harold's election and consecration, followed by a group of Anglo-Saxons pointing at the sky towards a 'star'. Bayeux Tapestry, scenes 30-32. Image: Bayeux Museum.

the tapestry designer is prepared to reinterpret history for a new purpose.

Harold and William then travel to Bayeux (scene 22), where the tapestry says Harold 'swore a sacred oath' (scene 23) (Figure 4). The Norman sources would have us believe that by this oath Harold promised to support William becoming king upon Edward's death, albeit none of them place this in Bayeux.

Oath made, Harold crosses the Channel (scene 24) back to England, where he is seemingly admonished by Edward (scene 25). This appears to follow the early twelfth-century account of Eadmer,[31] who says that the King reprimanded Harold for going to see William (see more below). The tapestry then reverses scenes again to connect the death of Edward (scenes 27-28) (Figure 5) with the election and coronation of Harold (scenes 29-31), as set out previously. Here, King Edward is illustrated almost touching hands with Harold, in line with the account of the *Vitae Ædwardi Regis*,[32] commissioned by Queen Edith; the tapestry shows Edith at the foot of her husband's bed as he makes Harold protector of the kingdom. Before this (scene 26) is the funeral of Edward and his burial. After is shown the election of Harold as king (scene 29), his consecration (scene 30), the men witnessing the coronation (scene 31) and the sighting of the comet (scene 32). As stated previously, important is how the tapestry connects these events – the coronation and Halley's Comet – to create a version of history to suit a particular narrative.

---

31   Eadmer, 8.

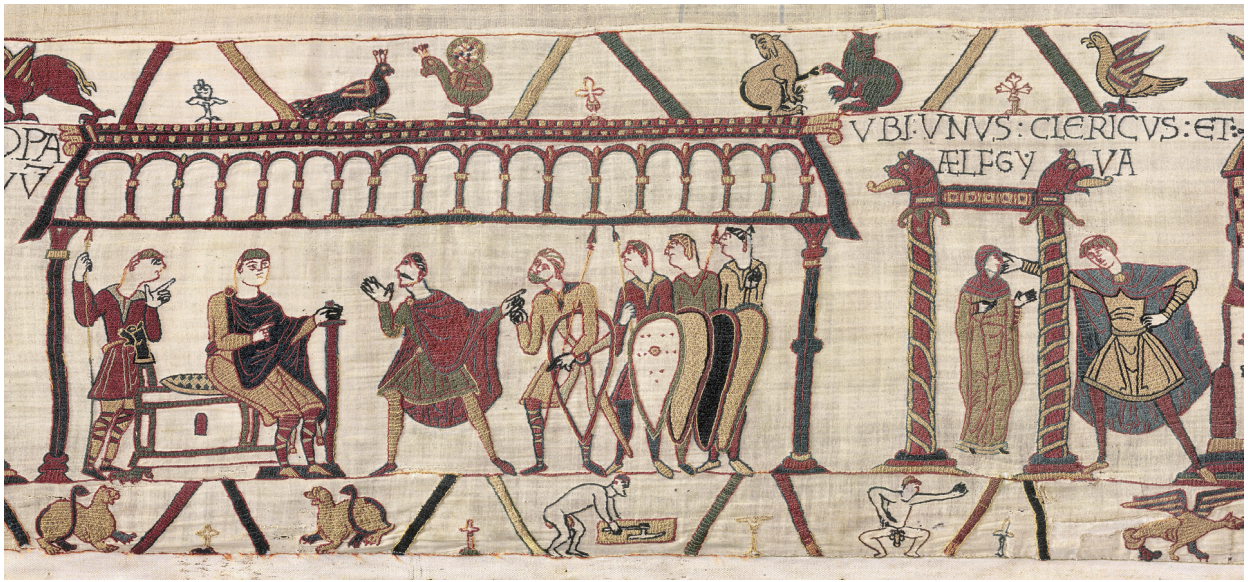

Figure 3: Harold and William talk, perhaps in the presence of a hostage, followed by the scene of the 'cleric and *Ælfgyva*'. Bayeux Tapestry, scenes 14-15. Image: Bayeux Museum.

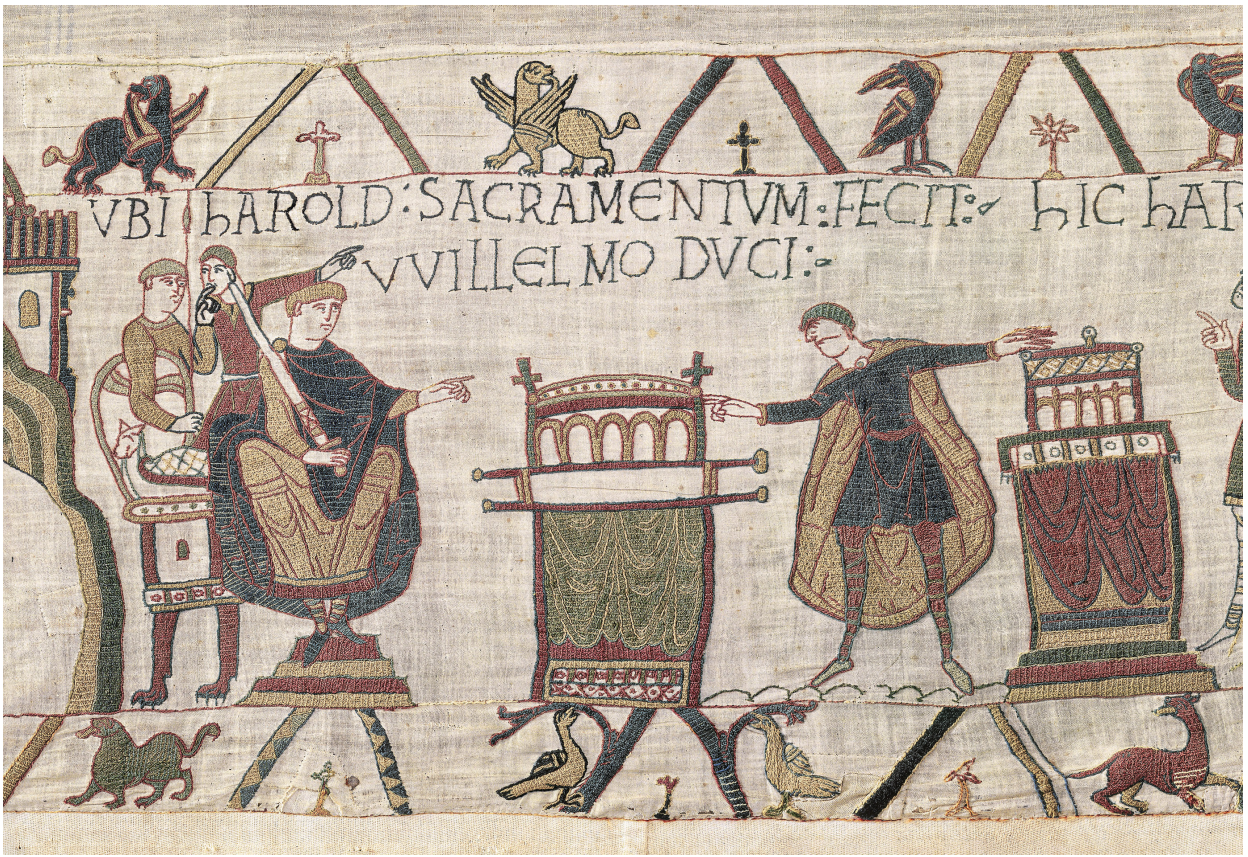

Figure 4: Harold making 'a sacred oath' at Bayeux. Bayeux Tapestry, scene 23. Image: Bayeux Museum.

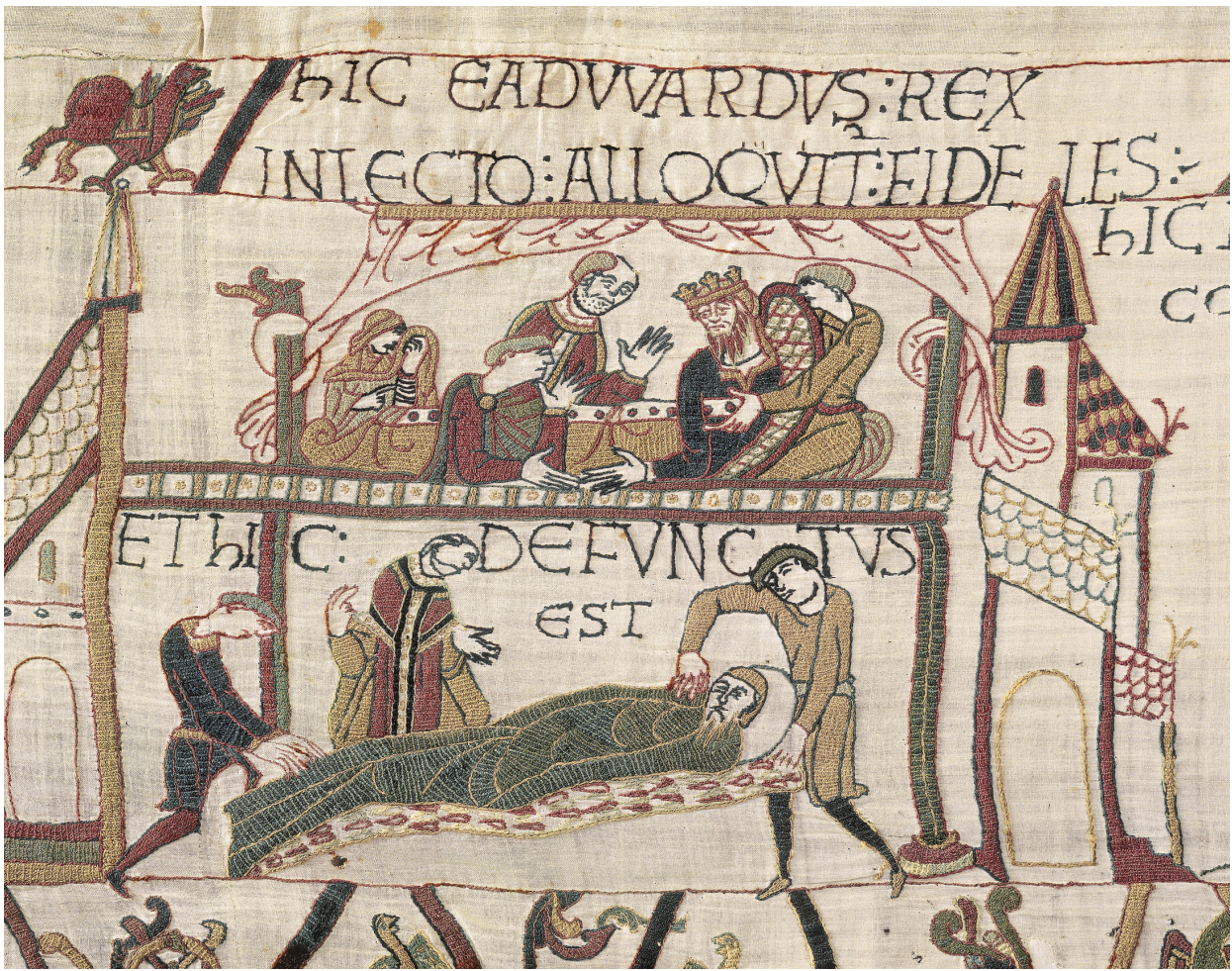

Figure 5: Edward's death, also (seemingly) showing the King making Harold protector of the kingdom. Bayeux Tapestry, scenes 28-29. Image: Bayeux Museum.

The 'coronation scene' is of additional interest because it depicts Archbishop Stigand. Although his appearance might be expected, as the Archbishop of Canterbury would have an important role in the consecration of a new king, Stigand is derided by William of Poitiers for holding the position at the expense of Robert of Jumièges, who was one of the 'foreigners' exiles in 1052 upon the return of the Godwin family in 1052. Importantly, Poitiers claims that Stigand's elevation to Canterbury was uncanonical, and therefore Harold's election was also unsound. It is important, therefore, that, in contrast, the Bayeux Tapestry recognises Harold's legitimacy as *rex Anglorum* (King of the English).

In 1066, Halley's Comet had its closest approach to the Sun on 23 March, approaching Earth to within 15 million km on 23 April.[33] The comet's arrival was also recorded in Chinese records on 2 and 19 April, being described as "bright or as large as the full moon" in the Korean *Goriosa*.³⁴ The Full Moon that year was on 13 April; on 19 April, it was only Last Quarter. There was also a partial lunar eclipse on 14 March; the annular eclipse on 28 April at 15:21 was not visible from England.³⁵ On 24 April, the Sóngh Sh records Halley's Comet as 'a broom' and 'visible in the morning at Mao'.³⁶ Its brightest tail that year (due to forward scattering) must have been around noon on 22 April.³⁷ Being as bright (or large) as the full moon,³⁸ the comet must have been visible in the afternoon and very apparent in the evening, assuming a clear sky. Therefore, Halley's Comet was not sighted before 24 April in England and possibly as late as 30 April,³⁹ several months after Harold's coronation (6 January). To explain this (apparent) error in the Bayeux Tapestry, which shows the comet soon after Harold's coronation, some scholars have argued that the tapestry might instead show another 'crown-wearing ceremony', perhaps at Easter 1066.[40] As will be shown, this argument makes some sense in the context of other primary sources that refer to the 1066 comet, even though these sometimes discuss its appearance in relation to other events. However, it is important to note that Easter (16 April) was at least one week before the comet appeared in the English skies. Even so, it seems likely that the tapestry designer purposefully placed Harold's coronation and Halley's Comet closer together than was the reality for political reasons with the benefit of hindsight.

---

32   *VÆ*, 79.

33   Kiang 1972.

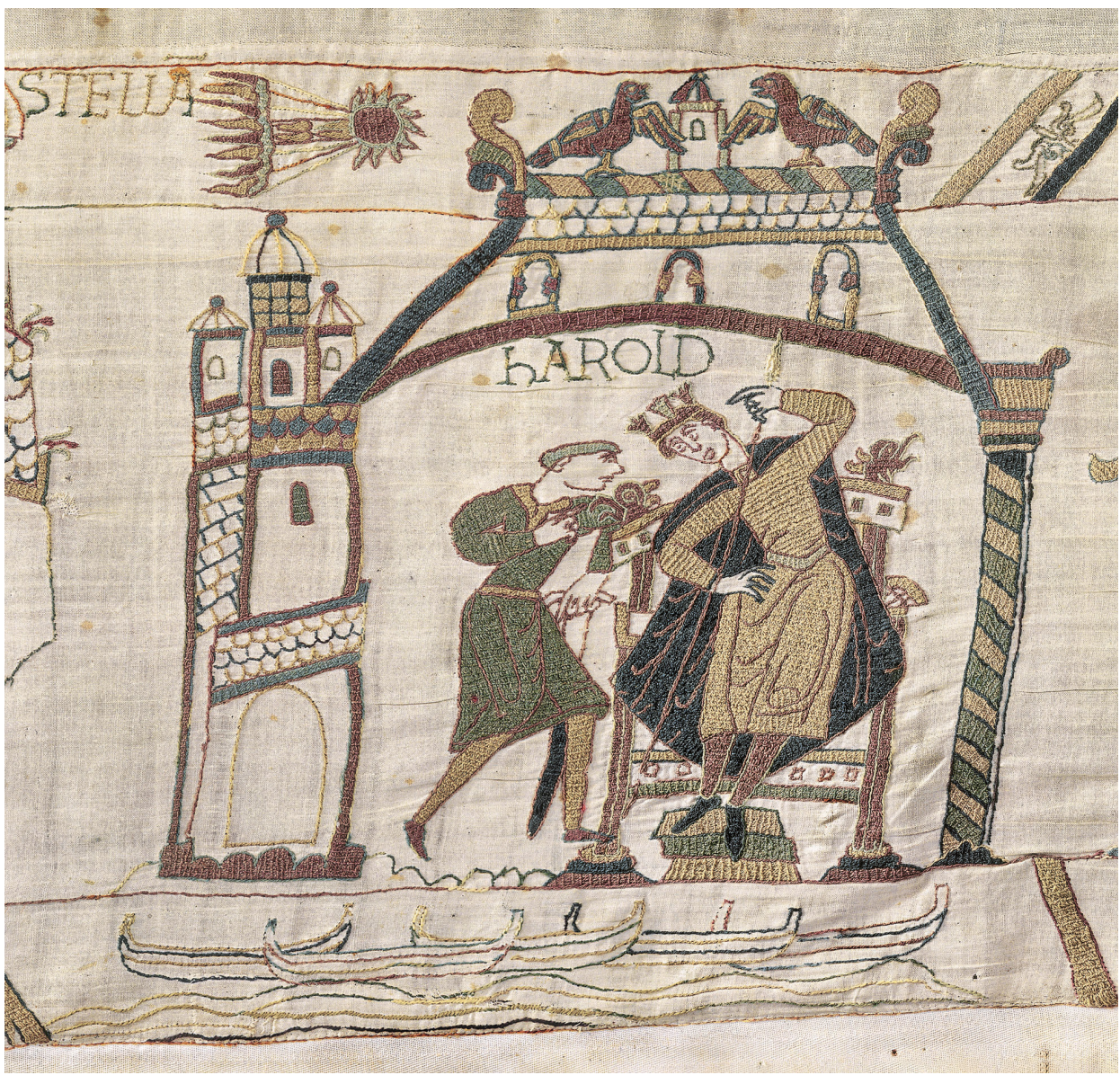

Figure 6: Harold receives some news. Bayeux Tapestry, scene 33. Image: Bayeux Museum.

---

34  Stephenson and Yau 1985, https://db.history.go.kr/goryeo/itemLevelKrList.do?itemId=kr.
35  https://moonblink.info/Eclipse/eclipse/1066_03_28_.
36  Lee *et al.* 2014.

37  Marcus & Seargent 1986.
38  Estimated to have an apparent magnitude of -12; Pingré 1783.
39  Vsekhsvyatskii 1964.

Harold is shown enthroned again soon after the comet (scene 33), but this time without his signature moustache (Figure 6). The 'ghostly' ships below him (in the tapestry's border) suggest a premonition of the invasions (from both Norway and Normandy) England was about to face, or may be suggestive of the English fleet that Harold ordered to defend the south coast.[41] The new king is clearly shown receiving a message. Perhaps it is the news that William is building a naval fleet to invade England or even the comet. However, it is important to note that those who first saw the 1066 comet interpreted it somewhat differently than how it has been recorded in the historical record. Although the threat of attack was a reality, the likelihood an invader could conquer the whole country seemed improbable. From this perspective, it is useful to consider Halley's Comet (and its import) in other contemporary sources, albeit all of these were written once Harold had been killed and England had been invaded and conquered, though the effects of the Conquest were still to be fully realised.

### Halley's Comet in contemporary sources

The Anglo-Saxon Chronicle is an essential primary source, particularly for the tenth and eleventh centuries. Initiated by King Alfred 'the Great' of Wessex in the late ninth century, it is a series of works offering a year-by-year account of English history. Each version (now named with a letter) provides a slightly different narrative of events, drawing out what was important from a local perspective. Versions C, D and E are the most instructive on the Norman Conquest period, but their tone is often terse.

'Version C' was at Abingdon, Oxfordshire, during the mid-eleventh century. After recording the date of Easter 1066, it says that "throughout all England, a sign such as men never saw before was seen in the heavens. Some men declared that it was the star comet, which some men called the 'haired star'; and it appeared first on the eve of the Greater Litany, that is on 24th April, and shone thus all the week". Intriguingly, given the 'ghostly ships' we see in the tapestry (see Figure 7), this version of the *Chronicle* connects the comet's appearance with the ravaging of the south coast by Harold's brother Tostig.

Earl Tostig is an interesting and complex individual. He appears to be a favourite of King Edward and Queen Edith, being her third eldest brother. Tostig became earl of Northumbria in 1055, thereafter making it his mission to rule the North more in line with the South, including thorough taxation;[42] whilst Edward was king of all England, the North had a level of autonomy. In contemporary sources, including the *Vitae Ædwardi Regis*,[43] Tostig is criticised for suppressing the Northumbrians "with the heavy yoke of his rule".[44] This led them to rebel, notably killing all Tostig's housecarls and taking his treasure. Harold necessarily sympathised with the rebels' demands, leading to his brother's banishment from England. Consequently, Tostig became vengeful.

The *Anglo-Saxon Chronicle* (C) says that "soon" after the comet's appearance "came Earl Tostig from beyond the sea into [the area around the Isle of] Wight, with as great a fleet as he could get, and there he was given both money and provisions". From there, he "did harm everywhere along the seacoast where he could get to until he came to Sandwich", Kent.[45] The *Chronicle* then describes how King Harold repelled him, with Tostig going on to raid "Lindsey" [Lincolnshire] before making for Scotland, where he was given safe conduct and provisions. Importantly, perhaps in line with the version of events told by the Bayeux Tapestry, the *Chronicle* places Tostig's raiding in the context of Harold anticipating a Norman invasion, saying that "when King Harold, who was in London, was informed that his brother Tostig had come to Sandwich, he gathered a greater ship-army and also land-army than any king in the land had ever gathered before, because he was told for certain that Earl William from Normandy… wanted to come here and win this land". After seeing off Tostig, Harold takes his fleet to the Isle of Wight, where it "lay there all the summer and the autumn" waiting for William.[46]

In the event, William's invasion fleet was delayed, and with it, his opportunity to invade England diminished. Harold was unable to maintain a naval fleet off the south coast indefinitely: "When it was the Nativity of St Mary [8 September], the men's provisions were gone, and no one could hold them there any longer. Then… the king rode inland, and the ships were sent to London".[47]

A very similar account of the comet is given in 'Version D' of the *Anglo-Saxon Chronicle*, probably composed at Worcester.[48] The comet does not appear in 'Version E', which was copied from another edition in the early twelfth century at Peterborough, showing how certain events mattered more to one chronicler than another. Importantly, the *Chronicle* accounts show that the appearance of Halley's Comet is less relevant in the context of the 1066 succession crisis than in the Bayeux Tapestry, instead helping to provide some chronological framework for Tostig's raiding activities and William's invasion.

---

40  Licence 2020, 175 and 182.
41  Musgrove & Lewis 2021, 210-211.
42  Barlow 2002, 83-88.
43  *VAE*, 50.
44  *ASC* (E), 190.
45  *ASC* (C), 194.
46  *ASC* (C), 195-196.
47  *ASC* (C), 196.
48  *ASC* (D), 195-197.

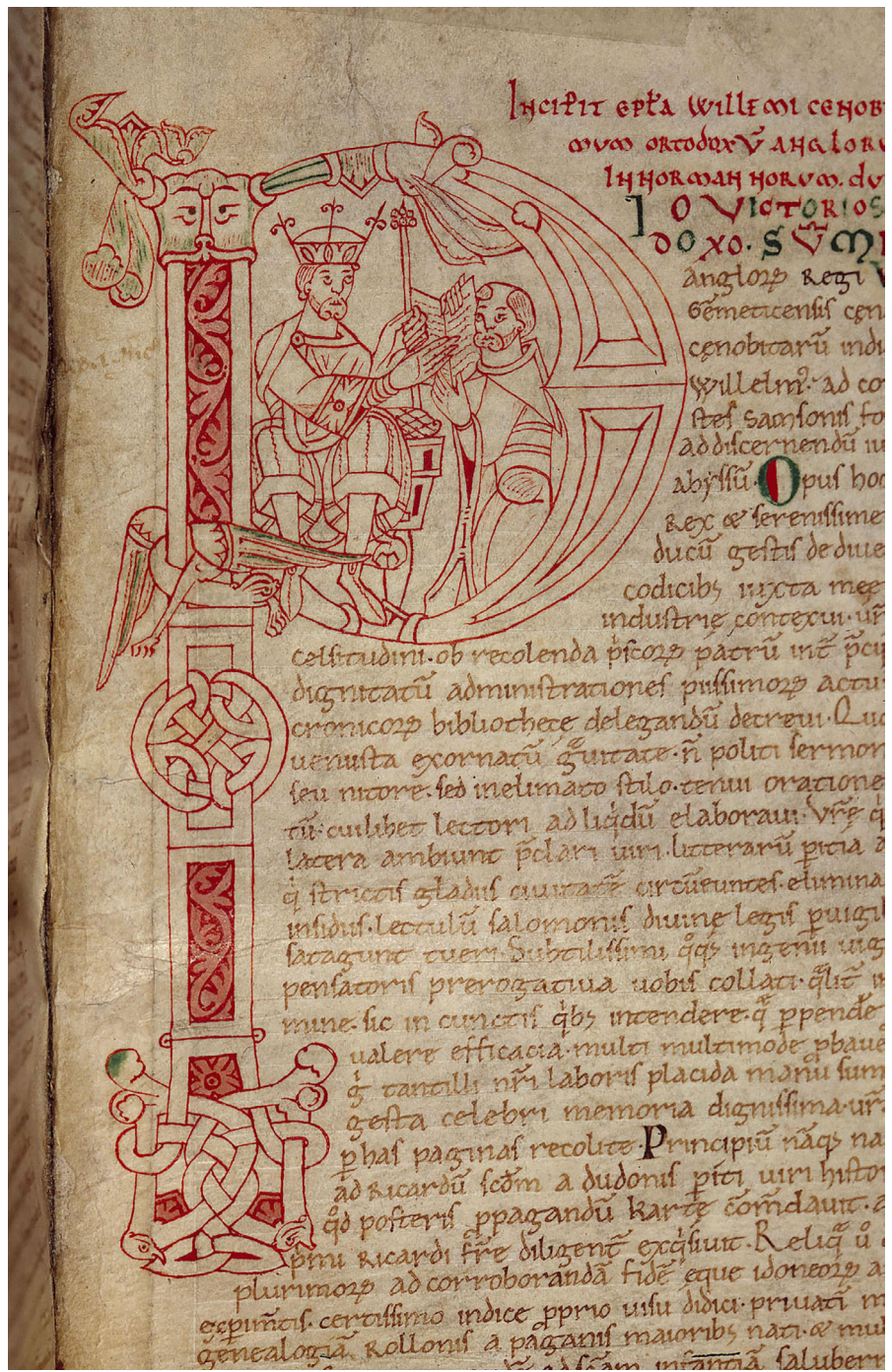

Figure 7: William of Jumièges presenting his *Gesta Normannorum Ducum* to William of Normandy, as King of England. Manuscript of c. 1070, Rouen, Bibliothèque Municipale. Image: Photo12/Archives Snark.

Neither is the 1066 comet recorded in the *Vita Ædwardi Regis*, the other main 'English' contemporary source. This was almost certainly written by a monk from the monastery of St Bertin at Saint-Omer, Flanders, in about 1067, perhaps by Folcard of St Bertin.[49] The *Vita* is dedicated to Queen Edith and probably commissioned by her. Intriguingly, this also omits Harold's visit to Normandy (in 1064 or 1065), which is key to the Norman argument in respect of Duke William's claim on England – though it does say Harold was "rather too generous with oaths (alas!)".[50]

---

49   Licence 2020, 282-297.
50   *VAE*, 53.

### The two Williams

The two main Norman sources for this period are written by William of Jumièges and William of Poitiers. The first William was a monk of the abbey of Jumièges, where he wrote the *Gesta Normannorum Ducum* ('Deeds of the Norman Dukes') in the 1050s and 1060s (Figure 8). This was extended, probably at the request of William of Normandy, once king of England, who visited the abbey in 1067. William of Poitiers' work, the *Gesta Guillelmi Ducis Normannorum et Regis Anglorum* ('Deeds of William, Duke of the Normans and King of the English'), was written in the 1070s, probably between 1071 and 1077. He seems to have been a knight in Duke William's service before becoming his chaplain.

William of Jumièges takes the view that Harold "seized Edward's kingdom... perjuring the fealty he had sworn"[51] to William in Normandy: this refers to Harold's visit in 1064-1065, which, according to this chronicler, resulted in him promising to help William become king upon Edward's death. William of Jumièges says that once Harold usurped the throne, Duke William dispatched messengers "urging him to renounce this act of folly", but Harold would not listen. It is after this that William of Jumièges mentions Halley's Comet, recording "at that time a star appeared in the north-west, its three-forked tail stretched far into the southern sky remaining visible for fifteen days". He then says that "it portended, as many said, a change in some kingdom".[52]

The fact that he only infers its connection to the English succession crisis of 1066 is intriguing, suggesting that only after the success of the Conquest were those recording such events putting two and two together. The comet first appeared in Normandy on 26 April, so two days later than in England, but it was probably not until late spring or the summer that William "moved into top gear" with his invasion plans.[53] Before then, Duke William needed to convince his followers that his plans to invade England had a reasonable chance of success. In this context, it is significant that William of Jumièges was writing soon after the invasion, so before the Conquest had fully played out, whereas William of Poitiers wrote some years later, benefiting much more from hindsight.

Indeed, William of Poitiers does not hold back in declaring the comet's significance regarding the succession crisis of 1066. His style evokes that of classical authors, connecting Duke William with great military leaders of the past. His rhetoric links Harold's taking the crown with his demise, saying: "Behold, you will not rejoice in the crown which you seized perfidiously, nor will you sit on the throne which you proudly mounted. Your end proves by what right you were raised through the death-bed gift of Edward. The comet, terror of kings, which burned soon after your elevation, foretold your doom".[54] Importantly then, even though Halley's Comet arrived over three months after Harold's coronation, William of Poitiers sees its coming as a retrospective premonition!

### Another Norman view

The *Carmen de Hastingae Proelio*, attributed to Bishop Guy of Amiens, is another account in the Norman tradition. It is a complex source, with its authenticity once being questioned. It is generally agreed that it was composed in about 1068, so, like William of Jumièges, it is more-or-less contemporary with the events it discusses. Since its poetic prose is not straightforward, it is often passed over by scholars. However, importantly for us, it does mention Halley's Comet. In contrast to the accounts of both Williams (of Jumièges and Poitiers), the *Carmen* suggests that the comet appeared in the sky when the Normans landed in England (28 September): "When you [William] left the sea behind and seized a safe beachhead, the third hour of the day was reaching over the earth, and a blazing comet with outstretched tail informed the English of their destined ruin". Of course, this could not have been true, but (again) it shows how contemporary sources distorted reality to connect certain events. In this sense, it is also noteworthy that the *Carmen* immediately thereafter discusses Harold's "treacherous... destruction of his brother [at the Battle of Stamford Bridge] in the remotest reaches of the realm". As such, it connects Halley's Comet with both invasions of England in 1066.

Following his raiding of England's coast and welcome by the King of the Scots, Tostig joined King Harald 'Hardrada' of Norway in invading Northumbria. Perhaps the aim was to install Harald as king in the North and reinstate Tostig as earl of Northumbria. Their expedition started well but ultimately failed. On 20 September, the Norwegians and Tostig defeated an English army led by Earls Edwin and Morcar at Fulford Gate, south of York. They then seized York itself before moving east and camping at Stamford Bridge. Here, on 25 September, the Norwegians were surprised by King Harold, with their army being destroyed; both Harald and Tostig were killed. Harold's victory was so comprehensive that the *Anglo-Saxon Chronicle* (D) records that only 24 ships were needed to take the Norwegians home.[55]

It seems that over time, the comet story developed from being a chronological marker, as witnessed in some of the earliest accounts, to being more certainly connected with Harold's demise and William's victory. In this sense, perhaps later chroniclers, who had even more time to reflect on the appearance of Halley's Comet and the succession crisis of 1066, can offer more insights.

---

51   Jumièges, 160-161.
52   Jumièges, 163. See also Seargent 2009.
53   Bates 2016, 217-220.
54   Poitiers, 140-144.
55   *ASC* (D), 199.

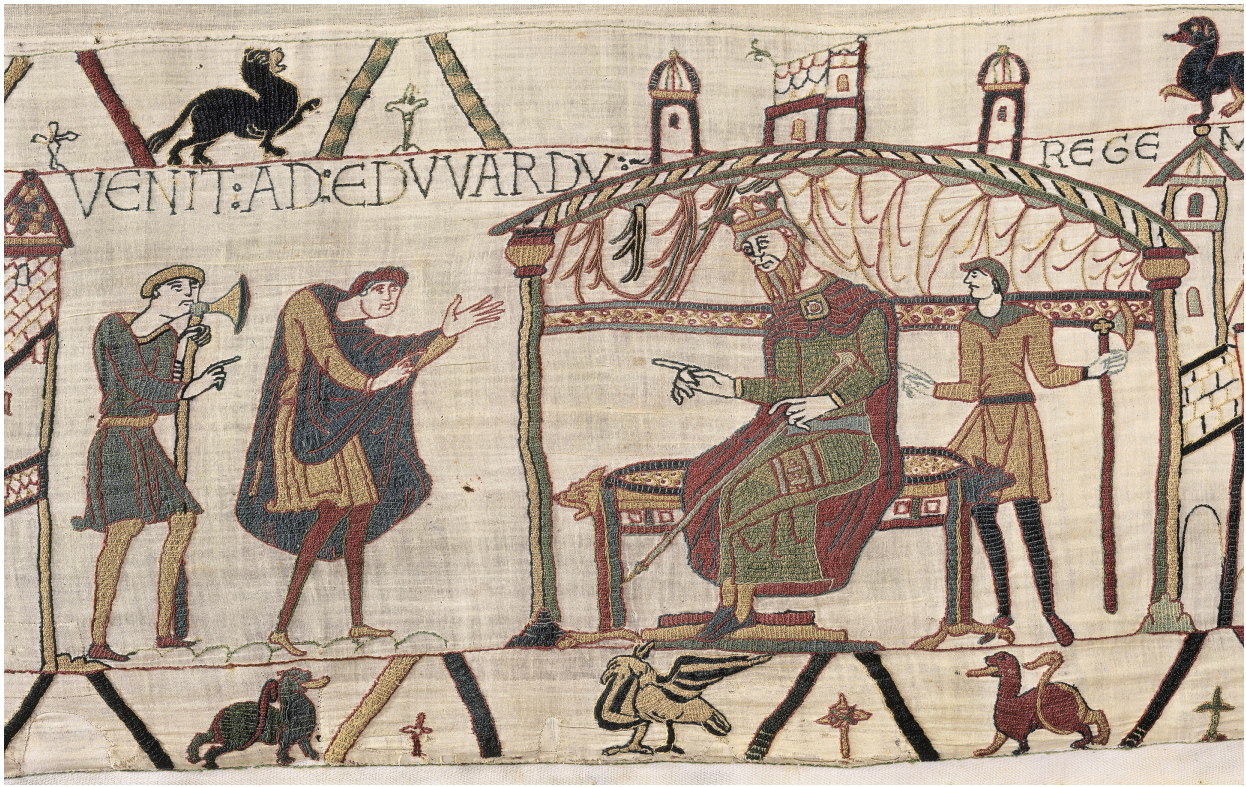

Figure 8: Harold seemingly grovelling before Edward. Bayeux Tapestry, scene 25. Image: Bayeux Museum.

**Anglo-Norman accounts**

Providing a view from the Anglo-Norman period is the English-born monk Eadmer. His *Historia Novorum in Anglia* ('History of recent events in England'), written between about 1093 and 1119, principally concerns ecclesiastical affairs but also reflects on other matters. Eadmer can be disparaging of the 'new regime' and its leaders but is also critical of Harold's actions in the lead-up to 1066, especially in terms of ignoring King Edward's advice. As has already been noted, Eadmer believed that Harold had travelled to Normandy to retrieve family members held hostage there. He records Edward's concerns about the trip, with the King saying to Harold: "I will have no part in this; but, not to give the impression of wishing to hinder you, I give you leave to go where you will and see what you can do [to bring the hostages back]. But I have a presentiment that you will only succeed in bringing misfortune upon the whole Kingdom and discredit upon yourself".[56] Edward is right, of course!

Upon Harold's return to England, Eadmer says he explains to Edward that he had agreed to support William becoming king and, worse, had made that promise upon holy relics.[57] Edward is furious, as seemingly shown in the Bayeux Tapestry (scene 25, Figure 8). Whether by accident or design, Harold certainly seems to grovel before the king, but are we reading too much into this depiction in the light of later sources? An alternative reading is that Harold is shown bowing (in courtesy) as he meets the king. Sadly, its Latin inscription – "Here Duke Harold returned to England and came to King Edward" – gives little away.

Eadmer does not mention the comet, but it is discussed by another near contemporary of his, John of Worcester. John was an English monk based at Worcester Priory. His *Chronicon ex chronicis* ('Chronicle of chronicles') gives a history of the Kings of England up to 1141, though most of it was written up to 1117 or 1118, more than half a century after 1066. John of Worcester takes a more sympathetic view of Harold than Eadmer, saying that on becoming King, he soon "destroyed iniquitous laws, and set about establishing just ones; becoming patron of churches and monasteries…"; indeed, his recording of Harold's good deeds goes on and on.[58]

Following the earlier English sources, John of Worcester records the comet in connection with the raids on the south coast by Earl Tostig. He says: "On 24 April, a comet was seen, not only in England but also, so they say, throughout the whole world, blazing for seven days in great splendour".[59] Shortly afterwards, Tostig returned from Flanders to the

---

56    Eadmer, 6.
57    Eadmer, 8.

58    Worcester, 600-601.
59    Worcester, 600-601.

Isle of Wight before raiding the south and east coasts on his way north to Scotland. Nothing more is said of the comet, and neither does John of Worcester's matter-of-fact account dwell on its meaning. Perhaps even his comment that it was seen "all over the world" somewhat dismisses its relevance regarding events in England.

**Later Chroniclers**

William of Malmesbury is considered one of the foremost historians of the twelfth century. Born towards the end of the eleventh century, he was a product of the Norman Conquest, with his father being Norman and his mother English. William's adult life was spent as a monk at Malmesbury Abbey in Wiltshire, where he wrote many works. Most significant in terms of Halley's Comet is his *Gesta Regnum Anglorum* ('Deeds of the English Kings), which he completed in 1125; his *Historia Novella* provides a sequel, up to 1142.

William of Malmesbury mentions the comet of 1066 when recounting the deeds of an aged monk of his monastery named Æthelmær, who wrote on astrology. Influenced by the Greek legend of Daedalus, Æthelmær also attempted to fly – by fixing wings to his hands and feet! Unsurprisingly, like that of Daedalus' son, Icarus, Æthelmær's flight ended in disaster – with him breaking his legs and becoming "crippled". Stoically, Æthelmær recounts (through William of Malmesbury) that the reason for his failure was "that he forgot to fit a tail onto his hinder parts".[60]

Æthelmær had apparently also seen Halley's Comet in 989 and recalled it as a bad omen. This is possible, though Halley's Comet then was 0.3 magnitudes dimmer than during the 1066 passage, having its closest approach to within 58 million kilometers of the earth on 20 August 989 after it passed the sun on 9 September.[61] William of Malmesbury says Æthelmær crouched "in terror at the sight of the gleaming star" of 1066, calling out to it: "You've come, have you?... You source of tears to many mothers. It is long since I saw you; but as I see you now, you are much more terrible, for I see you brandishing the downfall of my mother-country".[62] Whether or not Æthelmær really saw Halley's Comet twice (77 years apart), its significance in 1066 was clear to William of Malmesbury.

Intriguingly, William of Malmesbury not only recounts this tale in the context of King Edward's failing health but also within a wider geopolitical context. He says that preceding the comet's appearance, "three popes, Victor [II, reigned 1055-7], Stephen [IX, reigned 1057-8], and Nicholas [II, reigned 1058-61], weakened the vigour of the Holy See by their successive deaths", that the Holy Roman Emperor Henry (III, reigned 1046-56) died and was succeeded by his son, another Henry (IV), "who inflicted much oppression… by his folly and wickedness", and in "that same year Henry [I], king of the French" (reigned 1031-60), described as "an active and skilful soldier", was killed by poison.[63] So, like some Norman chroniclers, he connects Halley's Comet with the English succession crisis of 1066 but also, more generally, as a portent of doom.

Also important for understanding the Norman Conquest period is the chronicler Orderic Vitalis, who (like William of Malmesbury) also had a French father and an English mother. He was born in Atcham, Shropshire, and schooled to be a monk locally at Shrewsbury, before going to Saint Evroul, Normandy. Orderic Vitalis' first literary effort was a continuation and revision of William of Jumièges' *Gesta Normannorum Ducum*, before turning his hand to the *Historia Æcclesiastica* ('Ecclesiastical History'). This was written in c. 1136-1141, providing a retrospective on Harold becoming king. The influence of William of Jumièges is clear in his account of the comet, upon which he recounts: "In the year of Our Lord 1066… during April, a star known as a comet appeared in the north-west and remained visible for almost fifteen days. Learned astrologers who investigate the secrets of natural science declared that this portended the transfer of a kingdom". He then goes on to add thoughts on its meaning: "Indeed, Edward king of the English… had died shortly before; and Harold son of Earl Godwin had usurped the kingdom of England…".[64] For Orderic Vitalis, then, Halley's Comet was undoubtedly connected with the English succession crisis of 1066.

Orderic Vitalis generally takes a balanced view of the events leading to the Norman Conquest in 1066. But on the English succession, he says: "The truth was that Edward had declared his intention of transmitting the whole kingdom of England to his kinsman William Duke of Normandy, first through Robert archbishop of Canterbury and afterwards through the same Harold, and had with the consent of the English made him heir to all his rights. Moreover, Harold himself had taken an oath of fealty to Duke William":[65] this oath he places at Rouen rather than Bayeux (as in the tapestry).

By connecting the comet to the English succession, Orderic Vitalis follows the narrative set out by the 'Norman apologists'. Intriguingly, however, given the focus of the earlier English sources on the deeds of Tostig, it is of note that Orderic Vitalis says that it was Duke William who sent Tostig to raid England in 1066.[66] He is the only authority to suggest this: Tostig was in Flanders from November 1065 until he crossed to England, so if this was the case, he had little time to meet with William directly.

---

60  Malmesbury, 414-415.
61  Kiang 1972.
62  Malmesbury, 412-413.

63  Malmesbury, 412-413.
64  Vitalis, 134-135.
65  Vitalis, 134-135.
66  Vitalis, 142-143.

### Other comets around the year 1000

These accounts in English, Norman and Anglo-Norman sources inevitably raise the question of whether such celestial happenings – comets in particular – had always been seen as portents of doom in England. Or was the 1066 comet something different?

The *Anglo-Saxon Chronicle* (D)[67] for 905 records the death of Æthelwold, cousin to King Edward 'the Elder', following the appearance of a comet on 20 October. Rather than Halley's Comet, which has its closest approach to Earth much later (15 July 912), this could have been comet C/905 K1; it was sighted on 18 May and had its closest approach to Earth on the 25 April at a distance of 31 million km.[68] It is also perhaps the same comet as recorded by John of Worcester, who says that in 906 [sic] "a comet was seen",[69] except that comet C/905 K1 must have already faded by that time.

Edward succeeded his father, Alfred 'the Great', in 899 – a succession disputed by Æthelwold because he was the son of King Æthelred, Alfred's elder brother and predecessor.[70] At this time, much of England was occupied by the Danes. Æthelwold sought an alliance with them and subsequently was accepted as king in Danish-held Northumbria and (latterly) Essex.[71] As their king, Æthelwold led the East Anglian Danes against Edward's lands in Wessex and Mercia, returning home laden with booty.[72] In response, Edward raided East Anglia. But when Edward ordered his army home, both the *Anglo-Saxon Chronicle* and John of Worcester recount that part of it – "the inhabitants of Kent" – refused.[73] This led to a bloody battle between the English and Danes, resulting in Æthelwold being killed.[74] As such, as with Harold's death at the Battle of Hastings, might comet C/905 K1 be seen as a precursor to a perjurer's demise?

The *Anglo-Saxon Chronicle* also recounts events that followed the death of King Edgar 'the Peaceful' and the succession of his son Edward 'the Martyr'.[75] A poem in 'Version A' of the *Chronicle* says that following Edgar's death (on 8 July 975), "Up in the heavens appeared a star in the firmament which heroes, firm in spirit, prudent in mind, men learned in science, wise 'truth-bearers', widely called by the name comet".[76] As with the 1066 comet, this was interpreted as a bad omen, for the chronicler next records that God's "vengeance was widely known" as a "famine over the earth throughout the nation of men" – thus, intriguingly, considering its arrival within a wider geopolitical sphere as well. There are two hyperbolic comets recorded in 975 (X/975 G1 and P1), which had their perihelion passage (respectively) on 15 April and 3 August.[77] It seems unlikely, though, that any of these was seen in England, so maybe news of the comet was recorded from elsewhere.

Somewhat suddenly, the *Chronicle* (A) then says that "afterwards the Keeper of heavens, Governor of angels, improved it [and] gave back bliss to each of the island dwellers through the fruits of the earth".[78] This blip in England's fortunes is at first hard to understand since Edgar was well thought of by the chronicler, and nothing less was expected of his successor, Edward.

Both 'Versions D and E' of the *Chronicle* have similar words to each other on the matter: "Edward, Edgar's son, succeeded to the kingdom; and then immediately in harvest-time in that same year, the star comet [probably X/975 P1] appeared" and then "came a great famine and very manifold disturbances throughout the English race".[79] 'Version D' explains that because of Edward's youth, some leading figures, including Ælfhere of Mercia, "broke God's law" by impeding monastic rule and dissolving monasteries.[80] John of Worcester agrees, saying: "The state of the kingdom was thrown into confusion", resulting in Ælfhere expelling abbots and monks from the monasteries in favour of secular clerics (and their wives).[81] He then records that after Edward's consecration and anointment as king, "a comet was seen in the autumn".[82] It seems, then, that the kingship of a young man was blamed for the kingdom's fortunes, of which the 975 comet was retrospectively interpreted as a precursor.

The *Anglo-Saxon Chronicle* (E-F) records the death of Archbishop Sigeric of Canterbury following the appearance of a comet; a later hand to the Canterbury version (F) describes this as being "haired", a term also used for the 1066 comet.[83] Enigmatically, there is no such comet on record; the nearest hyperbolic sighting is comet X/998 D1, which passed the Sun in 998 but was surely not visible in 995. Even so, 'Version F' records that Ælfric of Wiltshire was chosen as Sigeric's successor on Easter Day (16 April), thus connecting his election with the comet.[84] The chronicler adds: "This Ælfric was a very [wise] man", hence, in contrast to comets hitherto described, heralding a positive turn of affairs. Further, we are told (again, in this later hand) that Ælfric set about reform at Canterbury, though it has been argued that the extent of Ælfric's reforms outlined in the *Chronicle* is fictitious, likely exaggerated by a later Canterbury-based chronicler for a local audience.[85]

---

67 *ASC* (D), 93
68 https://ssd.jpl.nasa.gov/sb/great_comets.html
69 Wocester, 360-361.
70 Foot 2012, 30.
71 Morris 2021, 252-255.
72 *ASC* (D), 93-95.
73 *ASC* (D), 93-95, Worcester, 358-359.
74 Worcester, 360-361.
75 *ASC* (A, D and E), 118-122, Morris 2021, 315-319.
76 *ASC* (A), 120.
77 https://web.archive.org/web/20160805213824/http://pds-smallbodies.astro.umd.edu/comet_data/comet.catalog
78 *ASC* (A), 120-122.
79 *ASC* (D and E), 121.
80 *ASC* (D and E), 121.
81 Worcester, 426-427.
82 Worcester, 428-429.
83 *ASC* (D-F), 128-129.
84 *ASC* (F), 128.

Unsurprisingly, then, 'Version E', attributed to Peterborough, paints a different story.[86] Here Ælfric's promotion is dislocated from Sigeric's death following the comet; it is the entry for 996 that records Ælfric succession.[87] As such, the hypothetical comet of 995 is firmly connected with Sigeric's death, not Ælfric's election. Intriguingly, Sigeric is of note for advising King Æthelred II to pay tribute to the Danes to stop their attacks on England.[88] Whilst this plan had short-term success, it eventually led to the debilitating taxation known as Danegeld. Within this context, the story of Ælfric reforming Canterbury in 'Version F' of the *Chronicle* following the appearance of the 995 comet makes more sense since it was a common narrative for the latter years of Æthelred's reign that the demise of religion brought the Danes to England. Perhaps, then, God's wrath for the people's sins is aptly represented by this celestial happening.[89] More significant, however, is whether chroniclers were adding to their text (in the way of hagiography) to give 'credence' to the events they record – i.e. a crisis needed a comet (or similar celestial happening) in its telling to be believed.

**After 1066**

Turning to sightings of comets after the Norman Conquest, the *Anglo-Saxon Chronicle* (E) recounts that "after Michaelmas" (4 October) 1097 "a strange star appeared, shining in the evening and soon going to rest".[90] "It was seen in the south-west, and the ray that stood from it shining south-east seemed to be very long, and appeared in this way well-nigh all the week. May men considered it was a comet". John of Worcester appears to record the same "star", though saying it was sighted earlier – "on 29 September 1097 for 15 days" – and that it was "in the shape of a cross", which is surely relevant.[91]

The sighted object may have been comet C/1097 T1, which made its closest approach to the Sun on 22 September,[92] and was conceivably visible from Earth from 29 September. It is not likely that the hyperbolic comet X/1097 X1 could have been seen several months before 6 December, when it was still on its way towards the sun; instead, this was likely visible sometime in January or February 1098.

The *Chronicle* records that the 1097 comet follows an expedition by William II 'Rufus' into Wales "with a great raiding army".[93] It is told that the king lost many men and horses before the Welsh turned against their king and chose another leader. Thereafter William returned to England. The *Chronicle* also says that "Soon after this, Anselm the archbishop of Canterbury... went across the sea because it seemed to him that in this nation little was done according to justice and according to his direction", suggesting that the chronicler connects the 1097 comet with Anselm's exile rather than William's expedition into Wales.[94] John of Worcester adds further credence to this hypothesis, directly linking the comet's sighting to disagreements between William and his Archbishop, saying, "there arose immediately dissension between the King and Anselm... because from the time he had become archbishop he had not been allowed to hold a synod nor to correct the evil practices which had multiplied in England".[95] As in the case of Harold's demise after the 1066 comet, Anselm's exile was sometime after the sighting of the 1097 comet: he left England for Flanders on 8 November 1097, and from there, he went to France before leaving for Rome to complain to the Pope. This is consistent with comet C/1097 T1 being visible around September 1097.

The following year (1098), this time "before Michaelmas", the *Chronicle* (E) records another celestial event in that "the heaven appeared as if it were burning well-nigh all the night".[96] It seems unlikely that the chronicler is recording the 1097 comet again. It could have been the hyperbolic comet X/1097 X1, which had its closest passage to the Sun on 6 December 1097 and may have been visible from Earth around November 1097 or in the first two months of 1098. It could also be possible that he recorded the *aurora borealis*.[97] Intriguingly, its occurrence does not coincide with obvious portents of change, though before its arrival, the *Chronicle* notes the deaths of various ecclesiastics, as well as the killing of Earl Hugh of Shrewsbury in Anglesey "by foreign Vikings". Another grisly event happened in the summer at Finchamptsted, Berkshire, when "a pool welled up blood", though the chronicler gives no view on what it meant.[98] So, like some of the other comets around the year 1000, their relevance remains enigmatic. But if it was the northern lights, no comet is responsible for these happenings.

---

85  Brooks 1996, 256-259.
86  *ASC* (E), 129-131.
87  *ASC* (E), 131.
88  Morris 2021, 329.
89  see also Lavelle 2002, 94.
90  *ASC* (E), 233.
91  Worcester, 84-87.
92  https://ssd.jpl.nasa.gov

93  *ASC* (E), 233.
94  see also Seargent 2009.
95  Worcester 2, 86-87; Barlow 1983, 374-376.
96  *ASC* (E), 234.
97  Many thanks to Annemarieke Willemsen for this suggestion.
98  *ASC* (E), 234.

**Final thoughts**

The *Chronicle of Melrose*, probably first written by monks at Melrose Abbey in 1173-1174, then extended, records that "a comet is a star which is not always visible, but which appears frequently upon the death of a king, or on the destruction of a kingdom. When it appears with a crown of shining rays, it portents the decease of a king, but if it has streaming hair and throws it off, as it were, then it betokens the ruin of the country".[99] It seems, then, that the 1066 comet was the latter type.

Historical texts show that comets might precede an array of happenings besides the death of a ruler or the ruin of a country, including the demise of other notable figures, famine, exile and other distress. In general, they were interpreted as portents of doom, but just as likely, they might be linked to changes of note without obvious disastrous ramifications for wider society. As noted, it is also likely that comet stories were embellished, even invented. The reality is that change is an inevitable fact of human life – 1066 not only claimed the life of Harold II but also of Edward 'the Confessor' and Harald 'Hardrada'.

By their very nature, such celestial events could only be interpreted retrospectively. Upon Harold's accession (January 1066), and even when Halley's Comet was in view across England (April), Harold's demise (and that of the kingdom) was by no means certain. In fact, right until 14 October, Harold had good reason to believe that he might be successful in fending off threats from abroad. In the event, Harold decimated one invading army (from Norway) and was narrowly defeated by the other (Normandy). Had he been successful, then maybe the 1066 comet would have been interpreted as a portent of the futile attempts of William of Normandy to take England by force, but alas (for the English), it was not to be…

**Acknowledgements**

In the Summer of 2020, Simon Portegies Zwart visited the Bayeux Tapestry with his family. Surprised by the lack of awareness about the importance of Halley's Comet in the tapestry display, he invited Michael Lewis to present at Leiden Observatory's general astrophysics colloquium. Following discussions with the Year 1000 team from the National Museum of Antiquities in Leiden present there, they collaborated on this paper.

**About the authors**

Michael Lewis is Head of Portable Antiquities & Treasure at the British Museum and Visiting Professor in Archaeology at the University of Reading and in Social Sciences and Humanities at the University of Helsinki. He has an interest in medieval metal small finds, particularly those associated with everyday life and religious practice, but is best known for his research on the Bayeux Tapestry; he is a member of the Bayeux Tapestry Scientific Committee that is advising Bayeux Museum on the display and interpretation of the Tapestry.
MLewis@britishmuseum.org

Simon Portegies Zwart is a professor of computational astrophysics at Leiden Observatory. He is a member of the Royal Holland Society of Sciences and Humanities, and founder of the commission C.1B on computational astrophysics of the International Astronomical Union. His interests covers computable and non-computable problems in astrophysics, and ranges across all scales, processes, and phenomena. To mediate this research he developed the Astrophysical Multipurpose Software Environment.
spz@strw.leidenuniv.nl

---

99   *Chronicle of Melrose*, 37 and 178.